\documentstyle[twoside,fleqn,espcrc2]{article}

\newcommand{\AmS}{{\protect\the\textfont2
  A\kern-.1667em\lower.5ex\hbox{M}\kern-.125emS}}

\newcommand{\dmin}{D_{min}}
\gdef\journal#1, #2, #3, 1#4#5#6{               
    {\sl #1~}{\bf #2} #3, 1#4#5#6}             
\def\prd{\journal Phys. Rev. D, }
\def\prl{\journal Phys. Rev. Lett., }
\def\cmp{\journal Comm. Math. Phys., }
\def\np{\journal Nucl. Phys., }
\def\pl{\journal Phys. Lett., }
\def\npproc{\journal Nucl. Phys. B (Proc. Suppl.), }

\title{Finite Size Effects on the QCD Spectrum Revisited\thanks{Presented by
S. Gottlieb}}

\author{MIMD Lattice Calculation (MILC) Collaboration ---
C.~Bernard,\address{Department of Physics,
Washington University, St.~Louis, MO 63130, USA}
T.~A.~DeGrand,\address{Physics Department,
University of Colorado, Boulder, CO 80309, USA}
C.~DeTar,\address{Physics Department,
University of Utah, Salt Lake City, UT  84112, USA}
Steven Gottlieb,\address{Department of Physics,
Indiana University, Bloomington, IN 47405, USA}\hskip -2.5pt$^,$\hskip-1pt
\address{Department of Physics, Brookhaven National Laboratory,
Upton, NY 11973, USA}
\thanks{BNL is supported under DOE Contract DE--AC02--76CH00016.}
A.~Krasnitz,$^{\rm d}$
R.~L.~Sugar,\address{Department of Physics,
University of California, Santa Barbara, CA 93106, USA}
and D.~Toussaint\address{Department of Physics,
University of Arizona, Tucson, AZ 85721, USA}
}

\begin{document}

\begin{abstract}
We have continued our study of finite size effects in the QCD spectrum
on lattices ranging in size from $8^3\times 24$ to $16^3\times24$.
We have increased our statistics for quark mass $am_q=0.025$ for the smallest
lattice size.  In addition, we have studied
quark mass 0.0125 for lattice sizes $12^3\times24$ and $16^3\times24$.
These lattice sizes correspond to a box 1.8--3.6 fm on a side when the
rho mass at zero quark mass is used to set the scale.
We discuss the nucleon to rho mass ratio at a smaller value of
$m_\pi/m_\rho$ than previously studied with two dynamical flavors.
\end{abstract}

\maketitle

\section{MOTIVATION}

For some time, it has been recognized that there can be substantial
finite size effects
in the hadron spectrum if the box size is too small.  This was seen very
early on in the quenched approximation\cite{Lipps83}.  In the first dynamical
fermion spectrum calculations
done by some of us, we  
used two different volumes in each
case to try to assess the magnitude of this effect\cite{Gottlieb88}.
More recently, there
have been calculations at weaker coupling over a wider range of lattice
volumes showing very dramatic finite volume effects\cite{Ukawa90}.
Our approach is to work with stronger coupling so that we can go to large
physical volumes.  We do this in the hope that by precisely evaluating the
finite volume effects where they are small, we can set the physical box
size for weaker coupling calculations where it would be very costly to
go to such large physical volumes.  That is, we want to go to a very large
physical volume and directly demonstrate that our largest volumes are more
than adequate.  We thus determine which physical volume is sufficient to
achieve a 1 or 2\% accuracy, and suggest that physical
volume as the minimum for weaker coupling calculations.  We would also like
to make contact with the analytic work of L\"uscher\cite{Luscher}.

Another advantage of stronger coupling is that we can go to smaller values
of $m_\pi/m_\rho$ than is currently
possible at weaker coupling.  This light quark
mass region is of interest because $m_N/m_\rho$ should decrease in the
chiral limit and because this is where we should start to see the effects
of rho decay.  It is well known that the $m_N/m_\rho$ ratio has always come out
too large in lattice calculations.  It is possible that only when we make
both the quark mass lighter and the lattice spacing smaller will we approach
the
observed value.  Approaching the chiral limit at accessible lattice
spacings strikes us as more relevant to the eventual extrapolation than
decreasing the lattice spacing at say $m_\pi/m_\rho= 0.5$.


\section{SPECTRUM CALCULATION}

The couplings that we use correspond to the finite temperature crossover for
$N_t=6$.  Since the smallest lattice we use has $N_s=8$, we have no worry
that we are suffering from ``spatial deconfinement.''
Our minimum lattice size is about 1.75 fm.  We use two quark masses,
$am_q=0.025$ and 0.0125, which we will refer to as the heavier
and lighter mass.  (For details, see Table 1.)  

\begin{table}[tb]
\setlength{\tabcolsep}{1.0pc}
\newlength{\digitwidth} \settowidth{\digitwidth}{\rm 0}
\catcode`?=\active \def?{\kern\digitwidth}
\caption{Summary of Staggered Spectrum Runs}
\label{tab:runs}
\begin{tabular}{lclr}
\hline
$am_q$ & $6/g^2$&size&length \\
\hline
0.025   & $5.445$ & $8^3\times 24$ & 1780  \\
*0.025   & $5.445$ & $8^3\times 24$ & 1554  \\
0.025   & $5.445$ & $10^3\times 24$ & 1392  \\
0.025   & $5.445$ & $12^3\times 24$ & 1036  \\
0.025   & $5.445$ & $16^3\times 24$ & 1556  \\
0.0125  & $5.415$ & $12^3\times 24$ & 974  \\
0.0125  & $5.415$ & $12^3\times 24$ & 282  \\
*0.0125  & $5.415$ & $16^3\times 24$ & 1220  \\
\hline
\end{tabular}
\end{table}

All of our masses come from fits with two particles.  For the pion channel,
we assume two pseudoscalars.  For all the other channels, we assume two
particles of opposite parity.  To see the finite size effects, it is
important for us to find the optimal set of fits.  For all of our fits, we
use the full covariance matrix with 20 time units blocked
together to calculate the covariance matrix.  We fit from $\dmin$ to the
center of the lattice for all particles but the nucleon.
Due to antiperiodic boundary conditions the nucleon propagator should vanish
on the center plane, so we ignore its value there.
For the heavier quark
mass, we have four lattice volumes, and we choose the fit according to the
combined confidence level of all four fits.
The rho combined
confidence level is 0.25, 0.20 and 0.12, for $\dmin=4$, 5 and 6,
respectively.
Last year we presented pion masses based on single
particle fits.  For $\dmin=7$ and 8, they both have combined confidence levels
of 0.27.  However, the two particle fits with $\dmin=2$ that we report here
have a combined confidence level of 0.48.  More details of the combined
confidence levels are found in Ref.~\cite{Milc92}, which contains
a complete table of masses.
For the heavier (lighter)
quark mass, we use $\dmin=4$(3) for all particles other than the pion.  For
the pion, we use $\dmin=2$ for both masses.

\section{FINITE SIZE EFFECTS}

For the heavier quark mass, we presented results last year at
Tsukuba\cite{Milc_tsukuba}.  By
increasing the statistics on our smallest lattice, we have been able to
increase the statistical significance of the difference in masses between
the smallest and largest volume.  For the lighter mass, we only have
simulations for the two larger volumes.
Here, we can only bound the size of the effect.
The effect should be larger with smaller quark mass, but
we don't have a comparable range of volume.

In Table 2, we    
compare our 1991 results
with the new fits and increased statistics.  The
pion mass is actually a little smaller for $N_s=8$ than before; however,
the errors have decreased, so the finite size effect is smaller but
statistically more significant.  Most of the effect seems to
occur between $N_s=8$ and 10.  In Fig.~1,   
we show the pion mass as a function of $N_s$.

\begin{table}[bt]
\setlength{\tabcolsep}{0.81pc}
\caption{Hadron Finite Size Effects}
\label{tab:pion}
\begin{tabular}{lll}
\hline
&1991 Result&1992 Result\\
\hline
$\pi(N_s=8)$& 0.4542(15) & 0.4529(7)\\
$\pi(N_s=16)$& 0.4483(6) & 0.4488(4)\\
Difference & 0.0059(16) & 0.0041(8)\\
Significance& 3.7$\sigma$ & 5.1$\sigma$\\
Percentage& 1.3\% & 0.9\% \\
\hline
$\rho(N_s=8)$& 0.947(18) & 0.949(12)\\
$\rho(N_s=16)$& 0.918(5) & 0.918(4)\\
Difference & 0.029(19) & 0.031(13)\\
Significance& 1.5$\sigma$ & 2.4$\sigma$\\
Percentage& 3.2\% & 3.4\% \\
\hline
$N(N_s=8)$& 1.482(55) & 1.456(22) \\
$N(N_s=16)$& 1.380(8) & 1.375(8) \\
Difference & 0.102(56)& 0.081(23)\\
Significance& 1.8$\sigma$ & 3.5$\sigma$\\
Percentage& 7.4\% & 5.9\% \\
\hline
\end{tabular}
\end{table}
\begin{figure}[tb]
\vspace{9.8cm}
\includegraphics{fig1.ps}
\caption{Pion mass as a function of lattice size}
\label{fig:pion}
\end{figure}
\setcounter{topnumber}{2}
\renewcommand{\topfraction}{0.9}
\begin{figure}[tb]
\vspace{5.4cm}
\includegraphics{fig2.ps}
\caption{Rho mass as a function of lattice size}
\label{fig:rho}
%
\vspace{5.8cm}
\includegraphics{fig3.ps}
\caption{Nucleon mass {\it vs}. $N_s$}
\label{fig:nucleon}
\end{figure}

Turning now to the rho, we see that there is an effect
slightly larger than three percent between the smallest and largest
lattice.  For the nucleon, we have the largest effect, almost six percent.
Because of the reduced errors, the significance of this effect has gone
from $1.8\sigma$ to $3.5\sigma$.  The masses are plotted as a function of
$N_s$ in Figs. 2 and 3.  
Results for the $\pi_2$ and $\rho_2$ may be found in Ref.~\cite{Milc92}.

How large must our box be to get within one or two percent of the infinite
volume limit?  It appears from our results that $N_s=12$ would suffice.  To
translate this into a physical size, we set the lattice spacing by
extrapolating the rho mass to zero quark mass and assuming that it has its
physical value.  In this way, we find that $m_\rho(m_q=0)=0.887(38)$
and the lattice spacing is 0.227 fm.  Our range $8 \le N_s\le 16$
corresponds to sizes of 1.8 to 3.6 fm, and we are suggesting that a box
size of 2.7 fm would be adequate for the quark masses studied here
(corresponding to $m_\pi/m_\rho\approx 0.4$).  Our calculations should be
compared with those done at $6/g^2=5.7$ where $m_\rho(m_q=0)= 0.340(16)$,
and the lattice spacing determined from the rho mass
is 0.087 fm.  Here $N_s$ ranges from 8 to 20 and
corresponds to sizes from 0.7 to 1.7 fm\cite{Ukawa90,Columbia}.
Recently, results have become
available for $N_s=32$ which corresponds to a size of 2.8 fm, but only
for $m_\pi/m_\rho=0.72$\cite{Schaffer92}.

\section{EDINBURGH PLOT}

Our Edinburgh plot is shown in Fig.~4.  
To calculate the error on
the ratio of hadron masses, we have simply added the individual percentage
errors.  This naive procedure probably overestimates the size of the error
since the hadrons' masses are all correlated.
Nevertheless, we have quite high precision, particularly for the largest
volume (octagon for $m_q=0.025$ and fancy square for $am_q=0.0125$).
Last year, we had some preliminary results for $am_q=0.0125$ on our smaller
lattice that indicated an improvement in $m_n/m_\rho$ as the quark mass is
decreased.  On our larger lattice, the nucleon mass has increased, and the
rho mass has decreased.  The mass ratio thus
increased by about 2 standard deviations when the volume was increased.
Unfortunately, this puts the nucleon to rho mass ratio back to about 1.5.
At this point, we must conclude that a light quark mass at this strong
coupling is not sufficient to cause the nucleon to rho mass ratio to
decrease toward its observed value.  It seems essential to decrease
the lattice spacing.  (See Ref.~\cite{Ukawa_rev}.)
\begin{figure}[tbh]
\vspace{5.9cm}
\includegraphics{fig4.ps}
\caption{Edinburgh plot for this calculation}
\label{fig:edinburgh}
\end{figure}

\vskip-12pt
\section{CONCLUSIONS}

We have explored finite size effects by looking at large lattices with high
statistics, though at a fairly strong coupling.
We see significant effects for the pi, rho and nucleon.
The box sizes used here ranged from about 1.8--3.6 fm.  Much of the
observed effect occurs between 1.8 and 2.3 fm.  A size of 2.7 fm seems
adequate for our higher mass to get results within 1 to 2 percent.
For the lighter quark mass, we really should explore smaller and larger
volumes.  At the 2\% level, we cannot see any finite volume effect between
$N_s=12$ and 16.

There seems to be little improvement in $m_N/m_\rho$ at this coupling as we
lower the quark mass.  Probably we need both weak coupling and light
quarks to reproduce the observed ratio.

It is important to study other quark masses and to try to make
quantitative
contact with analytic predictions.  Very high statistics will be required.
It is hoped that careful work at stronger coupling will set the physical
box size for the most aggressive calculations at weak coupling.

\section*{Acknowledgements}
This work was supported by both the U.S. Department of Energy and the National
Science Foundation.  Computations were done on Intel iPSC/860
supercomputers at the San Diego Supercomputer Center, the NASA Ames
Laboratory and the SSC Laboratory.


\end{document}